\documentclass{iopart}
\usepackage{epsfig}
\newcommand{\be}{\begin{equation}}
\newcommand{\bel}[1]{\begin{equation}\label{#1}}
\newcommand{\ee}{\end{equation}}
\newcommand{\bea}{\begin{eqnarray}}
\newcommand{\ba}{\begin{array}}
\newcommand{\eea}{\end{eqnarray}}
\newcommand{\ea}{\end{array}}

\begin{document}
\jl{1}
\comment[Comment on 'ECS of the Potts model']{Comment on
`Equilibrium crystal shape of the Potts model at the first-order
transition point'}
%                              %
\author{Sergei B. Rutkevich}
\address{Institut f\"ur Theoretische Physik, Universit\"at zu
K\"oln, D-50937 K\"oln, Germany, and
 Institute of Solid State and semiconductor Physics, P. Brovki str., 17,
 Minsk 220072, Belarus}
%\pacs{05.70.Ln, 45.70.Vn, 05.40.-a, 02.50.Ga}
%\submitted
\begin{abstract}
We comment on the article by Fujimoto (1997 {\it {J. Phys. A:
Math. Gen.}} {\bf 30} 3779) where the exact equilibrium crystal
shape (ECS) in the critical $Q$-state Potts model on the square
lattice was calculated, and its equivalence with ECS in the Ising
model was established. We confirm these results, giving their
alternative derivation applying the transformation properties of
the one-particle dispersion relation in the six-vertex model. It
is shown, that this dispersion relation is identical with that in
the Ising model on the square lattice.
\end{abstract}
%%%%%%%%%%%%%%%%%%%%%%%%%%%%%%%%%%%%%%%%%%%%%%%%%%%%%%%%%%%%%%%%%%%%%%%%
In paper \cite{Fuj} Fujimoto determined the equilibrium crystal
shape (ECS) in the $Q$-state Potts model on the square lattice at
the first order transition temperature. Fujimoto claimed
\cite{Fuj,Fuj1}, that ECS is universal for a wide class of models,
including the eight-vertex model \cite{Fuj1,Bax}, and the Ising
model on the square lattice \cite{Zia}. The origin of this
universality is still not well understood.

The subject of the this comment is to show, that the one-particle
spectrum in the six-vertex model (which is known to be equivalent
to the critical $Q$-state Potts model), and in the Ising model on
the square lattice are also the same. So, one can say, that the
universality of ECS in different square-lattice models reflects
the universality in the one-particle dispersion relation.

\begin{figure}
\begin{center}
\epsfbox{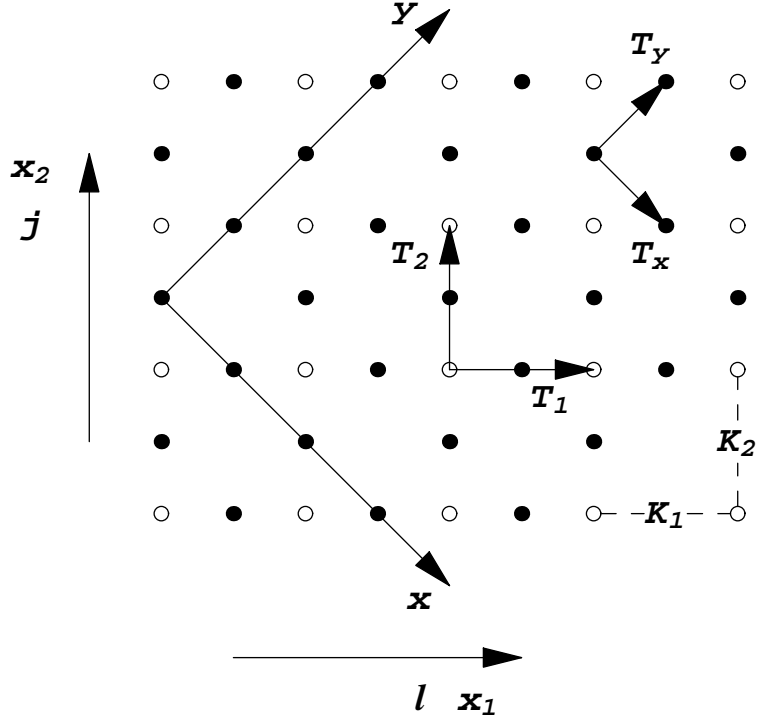}
\end{center}
\caption{\label{lat} The Potts model and the associated six-vertex
model square lattices. The Potts model sites and two-spin
interactions are shown by open circles, and by broken lines.
Indices $l$ and $j$ enumerate rows and columns of the Potts
lattice. The sites of the six-vertex model are shown by black
points.}
\end{figure}

Consider the $Q$-state Potts model ($Q>4$) on the square lattice,
shown in figure \ref{lat}. The sites of the Potts lattice are
depicted by open circles. The model Hamiltonian is given by
\begin{equation}
\beta \,E=-K_1\sum_{j,\,l}
\delta(\sigma_{j,l+1}-\sigma_{j,\,l})-K_2\sum_{j,\,l}
\delta(\sigma_{j+1,\,l}-\sigma_{j,\,l}).
 \label{ham}
\end{equation}
Here we use the same notations as in \cite{Fuj}, constants $K_1,\,
K_2$ obey the critical temperature condition \cite{Bax}: $(\exp
K_1 -1)(\exp K_2 -1)=Q$.  Let $\psi$ be a vector associated with a
horizontal row of the Potts lattice, $ \psi = \psi(...,
\sigma_1,\sigma_2,...,\sigma_l,...)$. Denote by $T_1$ and $T_2$
the two shift operators \cite{Fuj,Fuj1}, shown schematically in
figure \ref{lat}. Operator $T_2$ is the row-to-row transfer matrix
of the Potts model, and $T_1$ acts on the vector $\psi$ as
\begin{equation}
T_1 \, \psi(..., \sigma_1,\sigma_2,...,\sigma_{l},...)=\psi(...,
\sigma_2,\sigma_3,...,\sigma_{l+1},...).
 \label{sh}
\end{equation}

It is well known, that the critical Potts model is equivalent to
the six-vertex model \cite{Bax,Buf}, and to the interaction round
a face (IRF) model \cite{Fuj,KS}. Consider for definiteness the
associated six-vertex model. Its sites are shown by full circles
in figure \ref{lat}. One can define two shift operators $T_x$ and
$T_y$ corresponding to translations along the directions $x$, and
$y$. Note, that operator $T_y$ is the row-to-row transfer matrix
of the six-vertex model. As it is clear from figure \ref{lat},
these shift operators are related with $T_1,\,T_2$ by the
equations
\begin{equation}
T_1 \cong T_x \,T_y, \quad T_2 \cong T_x^{-1} \, T_y.
 \label{2sh}
\end{equation}

Consider operator $B(\theta)$, generating a one particle
excitation from the ground state of the six-vertex model. This
operator transforms under translations $T_x,\, T_y$ as
\begin{eqnarray}
T_x \, B(z) T_x^{-1} &=& \exp [-ip_x (z)]\, B(z), \label{transf}\\
 T_y \, B(z) T_y^{-1} &=&
 \exp [-ip_y(z)]\, B(z), \nonumber
\end{eqnarray}
where equations 
\begin{eqnarray}
\exp[\,-i p_x(z)] &=& \sqrt{k}\,\,{\rm sn}\Big(z-\frac{i\,
K'}{2}\Big), \label{dis}\\
\exp [-i p_y (z)] &\equiv& \exp[-
\omega (z)]= \sqrt{k} \,\,{\rm sn}\Big(z-\frac{2 i\, K
u}{\pi}\Big), \nonumber
\end{eqnarray}
define in the parametric form the dispersion relation
$\omega(p_x)$ of the one-particle excitations in the six-vertex
model. Here ${\rm sn}$ is the Jacobian sn function to the modulus
$k$, $0<k<1$, the quarter periods $K,\, K'$ satisfy $
 K'/K=2\,\lambda/\pi.$
Parameters $\lambda$ and $u$ are related with the original
parameters of the Potts model by \cite{Fuj}:
\begin{eqnarray}
  \sqrt{Q} =2 \cosh \lambda , && \\
 \frac{\exp(K_1)-1}{\sqrt{Q}} &=& \frac{\sinh[\frac{1}{2}\, \lambda -u]}
 {\sinh[\frac{1}{2}\, \lambda +u]}. \nonumber
\end{eqnarray}
The elliptic modulus $k$ can be explicitly written as
\begin{equation}
k =\Bigg[1-\prod_{n=1}^\infty \Big(
\frac{1-q^{2n-1}}{1+q^{2n-1}}\Big)^8\Bigg]^{1/2},  \label{k}
\end{equation}
where
$$
 q=\exp(-2\lambda)=\Bigg(\frac{\sqrt{Q}-\sqrt{Q-4}}{2}\Bigg)^2.
$$ 
The one-particle spectrum (\ref{dis}) can be obtained in the standard manner \cite{McCoy,Kor} from the Bethe ansatz solution of the six-vertex model \cite{Bax}. In the more general case of the eight-vertex model, such calculations have been done by Johnson, Krinsky and McCoy \cite{McCoy}. 
It should be noted, that equations (\ref{dis}) describe also
the one-particle excitation spectrum in the IRF  and the critical
Potts model \cite{Fuj,KS}. In the critical Potts model these
particles can be interpreted as the domain walls between an
ordered and the disordered phases \cite{Del}.

Let us transform the dispersion relation (\ref{dis}) as follows.
Consider transformation properties of the operator $B(z)$ with
respect to the translations along the rows and columns of the
original Potts lattice. From (\ref{transf}), (\ref{2sh}) one
obtains immediately
\begin{eqnarray}
T_1 \, B(z) T_1^{-1} &=& \exp [-ip_1 (z)]\, B(z), \label{trP}\\
 T_2 \, B(z) T_2^{-1} &=&
 \exp [-ip_2(z)]\, B(z), \nonumber
\end{eqnarray}
where
\begin{eqnarray}
 p_1 (z) &=& p_x (z)+p_y (z),\label{mom}\\
 p_2 (z) &=& -p_x (z)+p_y (z). \nonumber
\end{eqnarray}
By use of (\ref{dis}), and applying the well-known properties of
the Jacobian elliptic functions, one can exclude parameter $z$
from (\ref{mom}):
\begin{equation}
a_1 \cos p_1 +a_2 \cos p_2 =1, \label{dP}
\end{equation}
where
\begin{eqnarray}
 a_1 &=& [(1+ \rho^2)(1+k^2 \rho^2)]^{-1/2}, \label{a1}\\
 a_2 &=& k\,\rho^2 \,[(1+\rho^2)(1+k^2 \rho^2)]^{-1/2},  \label{a2} \\
\rho &=&-i \, {\rm {sn}} \bigg( \frac {2 i u  K}{\pi}+\frac {i
K'}{2}
  \bigg).  \label{rho}
\end{eqnarray}
Since $p_1$ and $\epsilon \equiv i p_2$ define the particle
quasi-momentum and energy, respectively, equations
(\ref{dP})-(\ref{a2}) give a compact form of the dispersion
relation for one-particle excitations in the critical Potts model.
The latter is exactly the same, as the excitation spectrum in the
Ising model on the square lattice first obtained by Onsager
\cite{Ons}. The only difference is in the meaning of parameters
$k$ and $\rho$. In the Potts model they are given by equations
(\ref{k}), (\ref{rho}), whereas in the Ising model we had
\begin{equation}
 k = [\sinh (2 K_1)\, \sinh(2 K_2)]^{-1},  \qquad
 \rho =\sinh(2 K_2).\label{Is}
\end{equation}

Substitution   $p_1\to i x_2, \quad p_2\to i x_1$ into the
dispersion relation (\ref{dP}) yields \cite{Acu,R}  the
equilibrium crystal shape (ECS)
\begin{equation}
a_1 \cosh x_2 +a_2 \cosh x_1 =1, \label{ECS}
\end{equation}
where $x_1,\, x_2$ denote the Cartesian coordinates in the
equilibrium crystal boundary. This shape is the same in the
critical Potts model \cite{Fuj}, and in the Ising model
\cite{Zia}, if the temperature $T<T_c$ and the anisotropy ratio in
the Ising model are determined by (\ref{Is}).

It is interesting to note, that the dispersion relation (\ref{dP}) is relevant also to the simple Gaussian model on the square lattice, defined by the Hamiltonian
\begin{equation}
\beta \,E_G =\frac{1}{2} \sum_{j,\,l}
\{ a \varphi_{j,\,l}^2+ c_1 ( \varphi_{j,l+1}-\varphi_{j,\,l})^2+ c_2 ( \varphi_{j+1,l}-\varphi_{j,\,l})^2 \},
 \label{hamG}
\end{equation}
where $\varphi_{j,\,l}$ is a real continuous order parameter. Really, the two-point correlation function in this model can be written as 
\begin{equation}
\fl \qquad
\langle  \varphi_{j,\,l} \,  \varphi_{0,\,0}\rangle =\frac{1}{a+2\,c_1+2 \, c_2} \int_{-\pi}^\pi \frac{d p_1 \; d p_2}{(2 \pi)^2} \; \frac{\exp(i p_1 \, l+i p_2 \, j)}{1-a_1 \cos p_1- a_2 \cos p_2}, 
\label{cor}
\end{equation}
where
$$
 a_1 =\frac{ 2 \, c_1}{a+2\,c_1+2 \, c_2}, \qquad  a_2 =\frac{ 2 \, c_2}{a+2\,c_1+2 \, c_2}.
$$
The dispersion relation is determined by the pole of the integrand in (\ref{cor}), i.e. by equation (\ref{dP}). Since the one-particle dispersion relation determines the angular dependence of the correlation length, the latter is the same in a wide class of square-lattice models including the Ising, six-vertex, critical Potts, and the simple Gaussian model (\ref{hamG}). It is likely, that the origin of this universality lies in the symmetry of the square lattice. If this is the case, one could extrapolate this result to higher dimensions, speculating, for example, that the angular dependence of the correlation length in the three-dimensional Ising model on the cubic lattice could be identical with that in the Gaussian model like (\ref{hamG}) on the same cubic lattice. 

\vspace{.3cm}
\noindent {\textbf{Acknowledgments.}} %\vspace{0.3cm}
%\noindent
I am thankful to Dietrich Stauffer for his steady interest to this
work, and for helpful discussions. I am grateful to Andreas
Schadschneider who drew my attention to the article by Fujimoto
\cite{Fuj}.

\noindent This work is supported by the Deutscher Akademischer
Austauschdienst (DAAD).
%{\bf Acknowledgments}:

%\bibliographystyle{unsrt}


\section*{References}
\begin{thebibliography}{99}
\bibitem{Fuj}
Fujimoto M 1997 J.\ Phys.\ A~{\bf 30}, 3779-93

\bibitem{Fuj1}
Fujimoto M 1992 J. Stat. Phys.~{\bf 67}, 123-53

\bibitem{Bax}
Baxter R J 1982 {\it Exactly Solved Models in Statistical
Mechanics} (London: Academic)

\bibitem{Zia}
Zia R K P and Avron J E 1982 Phys. Rev. B~{\bf 25}, 2042-5

\bibitem{Buf}
Buffenoir E and Wallon S 1993 J.\ Phys.\ A~{\bf 26}, 3045-62

\bibitem{KS}
Kl{\"u}mper A, Schadschneider A and Zitartz J  1989 Z.  Phys. B
~{\bf 76}, 247-332

\bibitem{McCoy}
Johnson J D, Krinsky S and McCoy B M 1973 Phys. Rev A~{\bf 8},
2526-47

\bibitem{Kor} Korepin V E, Bogoliubov N M, and Izergin A G 1993 {\it Quantum Inverse Scattering Method and Correlation Functions} (Cambridge: University Press)

\bibitem{Del} Delfino G and Cardy J 2000 Phys. Lett. B~{\bf 483}, 303-8

\bibitem{Ons}
Onsager L 1944 Phys. Rev.~{\bf 65}, 117

\bibitem{Acu}
Acutsu Y and Acutsu N 1990 Phys. Rev. Lett.~{\bf 64}, 1189-92

\bibitem{R}
Rutkevich S B 2001 J. Stat. Phys.~{\bf 104}, 589-608

\end{thebibliography}
\end{document}